# LC nanocomposites: induced optical singularities, managed nano/micro structure, and electrical conductivity

V.V. Ponevchinsky[a], A.I. Goncharuk[b], V.G. Denisenko[a], N.I. Lebovka[b], L.N. Lisetski[c], M.I. Nesterenko[c], V.D. Panikarskaya[c], M.S. Soskin[*a]

[a]Institute of Physics, NAS of Ukraine, 46 Nauki Prosp., Kiev, 03028, Ukraine; [b]Institute of Biocolloidal Chemistry, named after F. Ovcharenko, NAS of Ukraine, 42 Vernadskii Prosp., Kiev, 03142, Ukraine; [c]Institute of Scintillation Materials of STC "Institute for Single Crystals" of the NAS of Ukraine, 60 Lenin Ave., Kharkov, Ukraine.

## ABSTRACT

Microstructure, phase transitions, electrical conductivity, and optical and electrooptical properties of multiwalled carbon nanotubes (NTs), dispersed in the cholesteric liquid crystal (cholesteryl oleyl carbonate, COC), nematic 5CB and their mixtures, were studied in the temperature range between 255 K and 363 K. The relative concentration $X$=COC/(COC+5CB) was varied within 0.0-1.0. The concentration $C$ of NTs was varied within 0.01-5% wt. The value of $X$ affected agglomeration and stability of NTs inside COC+5CB. High-quality dispersion, exfoliation, and stabilization of the NTs were observed in COC solvent ("good" solvent). From the other side, the aggregation of NTs was very pronounced in nematic 5CB solvent ("bad" solvent). The dispersing quality of solvent influenced the percolation concentration $C_p$, corresponding to transition between the low conductive and high conductive states: e.g., percolation was observed at $C_p \approx 1\%$ and $C_p \approx 0.1\%$ for pure COC and 5CB, respectively. The effects of thermal pre-history on the heating-cooling hysteretic behavior of electrical conductivity were studied. The mechanism of dispersion of NTs in COC+5CB mixtures is discussed. Utilization of the mixtures of "good" and "bad" solvents allowed fine regulation of the dispersion, stability and electrical conductivity of LC+NTs composites. The mixtures of COC and 5CB were found to be promising for application as functional media with controllable useful chiral and electrophysical properties.

**Keywords**: nematic, cholesteric, 5CB, COC, multi-walled carbon nanotubes, nanocomposites, aggregates, electrical conductivity, optical singularities

## 1. INTRODUCTION

Presently, the liquid crystalline (LC) composites, filled with nano-scale colloidal particles, attract ever more and more interest[1]. It was demonstrated that nanomaterial dopants with highly anisotropic (rod- or disc-like) shape can affect and improve the distinctive photonic and electro-optic characteristics of LC used for optical device and display applications. Particularly interesting are LC composites based on chiral nematic liquid crystals (cholesterics). These materials exhibit selective reflection and giant optical activity that can easily be regulated by electric field and temperature. Doping of cholesteric liquid crystals (CLC) by nanoparticles can dramatically enhance their optical and opto-electronic characteristics. Introduction of ferroelectric particles in a CLC results in significant increase of the birefringence and dielectric anisotropy, as well as expansion of band reflection, and allows reduction of the driving voltage of switching between bistable textures [2]. Addition of $SiO_2$ nanoparticles to the mixture of 5CB (39.75%) and cholesterol oleyl carbonate (COC, 60.25%) substantially affects the helical structure of the system, and introduced disorder leads to decrease in the phase transition temperature and loss of the ability of selective reflection [3]. The composites with magnetic nanoparticles dispersed in a chiral nematic LC are of great interest due to the possibilities of making one-dimensional photonic crystals [3].

---

[*] marat.soskin@gmail.com; phone +38-044-525-55-63

Doping of LC by highly anisotropic carbon nanotubes (NTs) allows reduction of the response time and driving voltage, as well as suppressing of the parasitic back flow and image sticking typical for LC cells [4]. Also, remarkable electromechanical [5] and electro-optical [6] memory effects, as well as ultra-low percolation thresholds [7,8], were discovered. The previous experiments with NTs dispersed in cholesteric mixtures have demonstrated the impact of NTs on the selective reflection spectra [9,10], and it was suggested that NTs could destroy the translational order in the smectic phase of a CLC [9]. Increasing of the concentration of chiral additive (cholesterol nonanoate) in the nematic LC did also affect the stability of 0.01% NTs dispersion accelerating the aggregation and sedimentation of NTs [10]. The observed destabilization effect was explained by strong interactions of NTs with the helical structure of cholesteric LC and by the decrease of the helical pitch. The dielectric studies of the cholesteric LC (mixture of chiral additive ZLI-811 with nematic E7), filled by 0.5% of NTs, also have demonstrated the presence of interactions between NTs and the LC director [11]. The chiral hybrid composites, based on the mixture of CLC and NTs, may be promising for construction of a gas sensor with high dynamic range [12]. The functional ability of such chiral hybrid composites is determined by the nature of integration of NTs networks into the cholesteric structure that provides strong sensitivity of the optical and electrical properties of material to the external chemical and physical factors. However, the nature of such integration is still unexplored, and little is known about the properties of chiral hybrid composites on the basis of NTs and CLC.

This work is devoted to the study of electrical conductivity, microstructure, phase transitions and optical properties of multiwalled carbon nanotubes, dispersed in a cholesteric LC (cholesterol oleyl carbonate, COC), nematic 5CB and their mixtures in the temperature range between 255 K and 363 K.

## 2. MATERIALS AND METHODS

The cholesteric COC (cholesteryl oleyl carbonate) was obtained from Aldrich, USA. Its molecules have a rigid and flat structure, caused by the presence of the, so-called, steroid (cyclopentaneperhydrophenantrene) condensed rings. Under cooling, pure COC exhibits the isotropic(I) → cholesteric(Ch) transition at $T_{ICh} \approx 309$ K, the cholesteric(Ch) → smectic A (SmA) transition at $T_{ChSm} \approx 295$ K and the smectic A (SmA) → solid(C) transition at $T_{SmC} < 273$ K. The nematic 5CB, 4-pentyl-4'-cyanobiphenyl, of 99.5 % purity (Chemical Reagents Plant, Ukraine) has elongated molecules with rigid biphenyl core and a semiflexible alkyl chain. 5CB demonstrates the nematic-isotropic transition at $T_{NI} \sim 308 – 309$ K and the crystal-nematic transition (melting) at TCN = 295.5 K. Molecular structures of COC and 5CB are compared in Fig. 1.

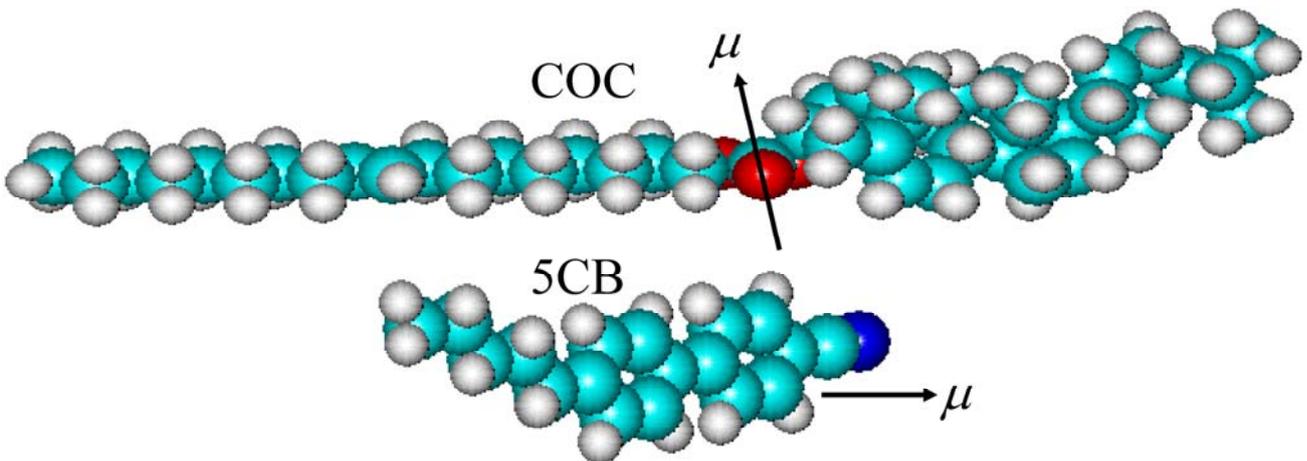

Figure 1. Comparison of the chemical structure of COC and 5CB molecules. Arrows show the direction of the dipole moments.

The multiwalled carbon nanotubes (NTs) were prepared from ethylene using the chemical vapor deposition method (TMSpetsmash Ltd., Ukraine) with FeAlMo as a catalyst [13]. They were further treated by alkaline and acidic solutions and washed by distilled water until reaching the distilled water pH value in the filtrate. The typical outer diameter of NTs was 25-30 nm, while their length ranged from 5 to 10 μm. The specific surface area of the NT powder, determined from $N_2$ adsorption, was $130 \pm 5$ m$^2$/g.

The COC+5CB+NTs suspensions were prepared by addition of the relevant quantities of NTs to the COC+5CB mixtures in isotropic (T=320 K) phase with subsequent 20-30 min sonication at the frequency of 22 kHz and the output power of 150 W using an ultrasonic disperser UZDN-2T (Ukrrospribor, Sumy, Ukraine). After sonication, the suspensions were rapidly cooled to the temperature ~350K. The weight concentration of MWCNTs was varied within 0-1.0 %.

The optical microscopy images were obtained also using OI-3 UHL 4.2 microscope (LOMO, Russia). The microscope CCD detector unit was interfaced with a digital CCD camera and a personal computer. A sandwich-type LC cell of 50 μm thickness was used and the samples were not oriented.

The high resolution Stokes-polarimetry images were obtained using a sandwich-type 20 μm thick LC cell with glass plates covered by rubbed polyimide coating. The optical Stokes-polarimetry is a very effective method for investigation of the detailed polarization structure of elliptically polarized optical fields with fine features typical for LC nanocomposites [14–17]. But an essential problem is to realize the Stokes-polarimetry using polarization microscope, because the coherence length of the light source has exceeded the thickness of an optical cell with an LC nanocomposite. These difficulties were overcome by using the light diode 3W LED Edixon S series EDER-SLC3-03 with75 Lm working in 620-630 nm as the light source. High quality polarization microscope POLAM L-213M (LOMO, Russia) was applied. Enlarged structures of the Stokes components were registered by CCD camera Lumenera LM-135M with 1392x1040 pixels resolution (1 pixel≈ 0.909 μm).

The phase transition temperatures were determined from the calorimetry experiments using a thermo-analytical system Mettler TA 3000 (Switzerland). The measurements were carried out in the temperature range of 278–363 K in the heating mode and were started from the super cooled smectic A phase (T=275 K). The mass of the samples was ~20 mg, and the scanning rate was 2 K/min, which ensured clear peaks on the obtained thermograms and enabled resolution of very small temperature shifts for materials under study. The heat flux and temperature calibration was done using the metallic indium standard at the same scanning rate as in experiments with the tested samples.

Cholesteric regular spiral structures form Bragg mirrors which reflect one-hand circular component and transmit opposite-hand circular component when the angle of incident light beam fulfills the Bragg condition [18]. Selective reflection spectra were measured using a Hitachi 330 (Japan) spectrophotometer within the 300-750 nm spectral range. A sandwich-type 50 μm thick LC cell was used. A sample was introduced between the cell walls using the capillary forces at the temperature of ~313 K, which corresponds to the isotropic phase of investigated medium. The measurements were done within the temperature range of 290-310 K in the heating mode (in order to prevent the supercooling effects). The temperature was stabilized by a flowing-water thermostat (±0.1 K).

The electrical conductivity measurements were carried out within 293–333 K, in the heating and cooling cycles, with the constant rate of scanning 2 K/min. The temperature was stabilized using a homemade thermoelectric cooler, based on Peltje elements (255-320 K), and was recorded by a Teflon-coated K-type thermocouple (±0.1 K), connected to the data logger thermometer centre 309 (JDC Electronic SA, Switzerland). The unoriented samples were used and the conductometric cell included two horizontal platinum electrodes of 14 mm in diameter, with 0.5 mm inter-electrode space. Before the measurements, the cell parts were washed in hexane and dried at 390 K. The electrical conductivity of the samples was measured by the inductance, capacitance, and resistance (LCR) meter 819 (Instek, 12 Hz–100 kHz). The measurements were done under the applied external voltage of 1 V and frequency of 500 Hz. This frequency was selected for avoidance of significant polarization effects on the electrodes.

## 3. RESULTS AND DISCUSSION

**Aggregation of nanotubes in COC+NTs composites**

Previous investigation of LC+NTs composites with different types of LC matrixes has evidenced that NTs form visually detectable individual aggregates even at small concentrations (≤0.01%) [17,19–21]. Further increase of *C* in the range of 0.05-0.1 % resulted in interconnection of isolated aggregates and, finally, in formation of a percolating network of NTs permeating into the LC matrix. The network of NTs radically increased electrical conductivity of LC composites. Visual inspection of COC+NTs composites revealed high level of NT homogenisation inside the LC matrix even at large concentrations of NTs.

Figure 2 presents microscope images of COC + NTs composites at different concentrations of NTs in the cholesteric phase (*a, T*=298 K) and isotropic phase (b, *T*=313 K). In the isotropic phase (*T*=313 K), the individual aggregates were not observed even at relatively large concentrations of NTs – these aggregates were becoming detectable at *C*≈0.3% and formed a continuous spanning network at *C*≈1.0%. In the cholesteric phase, the influence of NTs on the texture was even more evident at small concentration of NTs. Figure 3 shows an enlarged microscopic image of COC + NTs composites at *C*=0.3% and *T*=298 K. The observed different behaviour of NTs in the isotropic and cholesteric phases can be understood within the concept of formation of fractal NT aggregates in the orientationally ordered medium [8,19,20]. Interaction of nanotubes in the cholesteric phase leads to formation of individual NT aggregates, while the predominant tendency in the isotropic phase is formation of continuous NT networks (percolation).

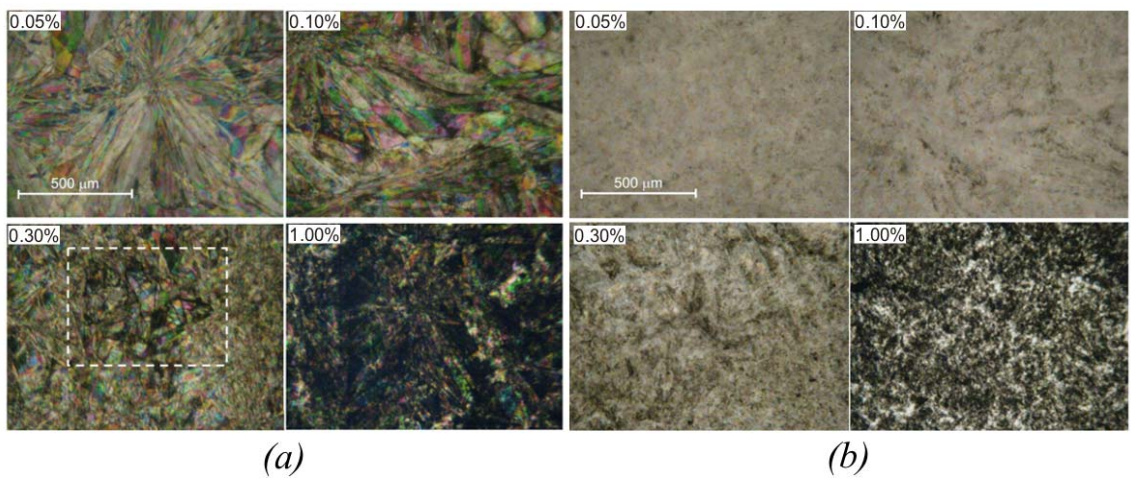

Figure 2. Microscope images of COC + NTs composites at different C: a) T=298 K (cholesteric phase), b) T=313 K (isotropic phase).

The presence of a dense network of defects was evident, and these defects could be ascribed to perturbation of the cholesteric layers, related to the presence of inclusions of small NT aggregates. The similar perturbation effects were previously observed for cholesteric LC filled by $SiO_2$ nanoparticles [22].

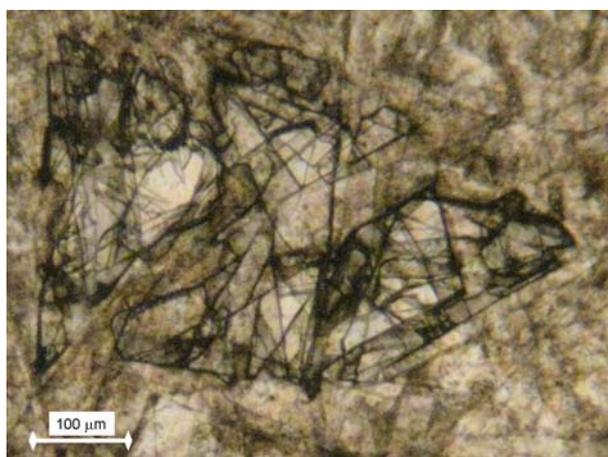

Figure 3. Enlarged portion of the microscope image of COC + NTs composites at C=0.3% and T=298 K (cholesteric phase).

The observed structure of NTs composite is nontrivial and requires supplementary investigations. Formation of long-ranged 100-500 μm long 1D ordered structures, stabilized by the relatively short (5-10 μm) individual NTs, may be supposed. As far as we know, such 1D ordering is observed for the first time. We can suppose that such structures appear due to self-organization of NTs inside the highly ordered spiral structure of a pure cholesteric.

**Optical properties and percolation transition in hybrid COC+5CB+NTs composites**

The typical examples of microstructure of COC+5CB+NTs composites are presented in Fig. 4. The patterns were rather homogeneous at large COC content ($X$= 0.7 – 1.0). However, more and more NTs clusters were becoming visible when COC content decreased. Below the level of $X \approx 0.5$, the disordered 50-100 μm clusters of nanotubes were observed inside the LC-matrix. The relatively thin (oily) lines can be seen also in the high-resolution Stokes-polarimetry images. Analysis has shown that polarization ellipses have very large eccentricity in the range of 0.1. The measurements of their azimuth along the horizontal line on the 200 μm height exhibit spikes in the points of crossing of the oily lines. One of the causes of their appearance are stresses in the LC medium provoked by the presence of NTs clusters. This phenomenon will be investigated in details in the future.

Note that both COC and 5CB have similar nematic phase diapasons. It is reasonable to expect that nematic phase diapasons of their mixtures are also similar and are in vicinity of the room temperature. It was justified by direct DCS investigations. So, the mixtures of "good" (COC) and "bad" (5CB) may be promising as nematic LC media allowing fine regulation of the aggregation structure of NTs. Increase of $X$ resulted in decrease of NTs aggregation; it means that the value of $X$ may affect the percolation threshold in the composites under consideration. We can expect that in pure COC the percolation transition will be observed at a rather high concentration of NTs and it may reflect two effects:

- High exfoliation and stabilization of the NTs in COC, and suppressing of aggregation ability of NTs.
- Formation of isolating solvate layers of COC molecules on the surface of individual NTs. These layers can inhibit the hopping transport of charge carriers between different NTs at small concentrations of NTs.

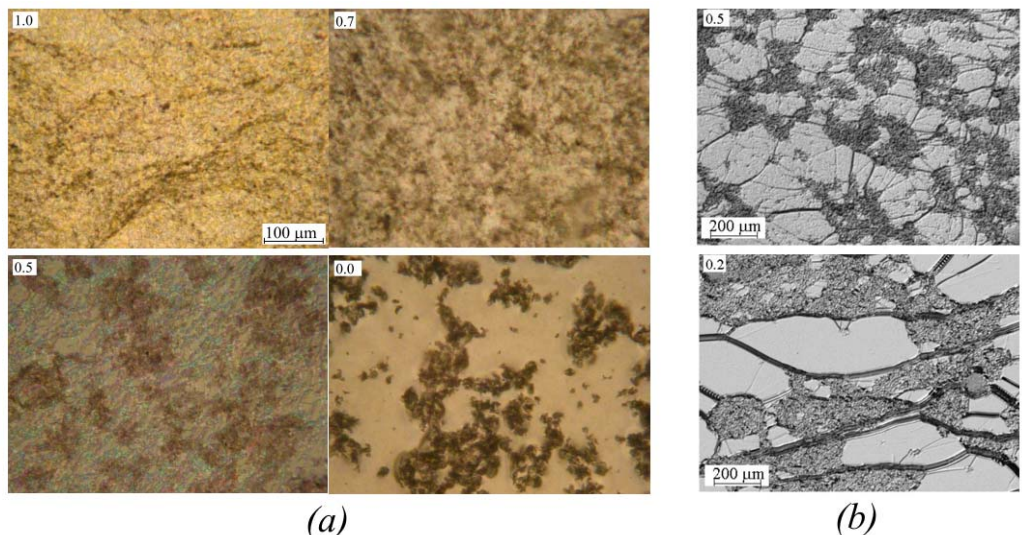

Figure 4. Micro-images of nanocomposites (*a*) and high resolution Stokes-polarimetry images (*b*). The relative concentration of COC $X$=COC/(COC+5CB) in the mixture containing 0.1% NTs, varies from 1.0 to 0.0.

Both effects are a consequence of strong interactions between NTs and COC molecules. So, COC is a relatively "good" solvent for NTs. From the other side, the NTs have demonstrated high tendency to aggregation in other LC solvents, such as 5CB, EBBA, BBBA [7,8,17,19–21,23], which are "bad" solvents for NTs as compared with COC.

**Phase transitions and selective reflection in COC+5Cmixtures**

Both COC and 5CB have the similar temperature ranges of the chiral nematic (cholesteric) and nematic phases that are close to the room temperature and almost similar temperatures of phase transition to the isotropic phase (≈309 K and 308.5 K, respectively). Introduction of 5CB into COC mainly affected the temperature of phase transition between the smectic and cholesteric phases $T_{ChSm}$.

Figure 5 shows dependence of $T_{ChSm}$ versus relative concentration $X$=COC/(COC+5CB) of COC in the mixture in the concentration range within 0.5 and 1.0. It is interesting to note that introduction of a small quantity of non-smectogenic 5CB into COC (i.e., at a relatively large $X$, $X$>0.8) resulted in increase in the thermal stability of the smectic-A phase. It evidences that 5CB molecules facilitate the tendency of COC molecules to translational smectic ordering, which reflects the effects of strong packing interactions between 5CB and COC molecules. The hydrocarbon chains of COC molecules have conformations far from the most extended in the cholesteric phase. Interactions with inflexible 5CB molecules can stabilize more elongated conformation of COC molecules and increase their tendency to smectic ordering. However, the decrease in COC concentration at $X<0.8$ resulted, as it could be expected, in decrease of $T_{ChSm}$, which was becoming lower than room temperature at $X<0.5$. The smectic ordering was disappearing completely below the certain critical value of $X$.

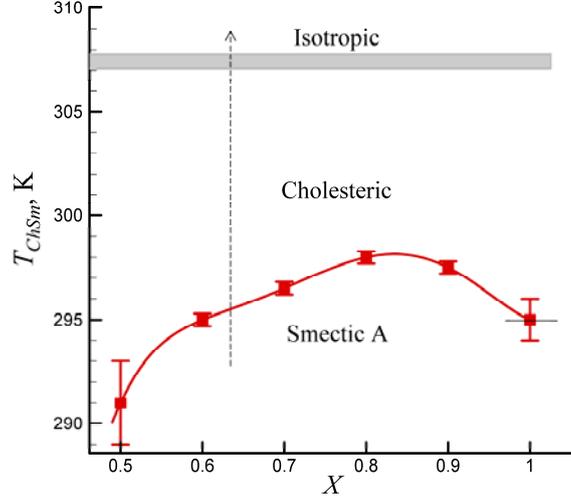

Figure 5. Phase diagram of COC+5CB mixtures at large values of the relative concentration $X$=COC/(COC+5CB). It was obtained by heating the supercooled smectic A phase.

Introduction of 5CB into COC affected also the helical pitch $\lambda$. Figure 6 presents reciprocal helical pitch $1/\lambda_m$ ($\lambda_m$ was estimated from the maximum of selective reflection) versus $X$ at $T$=306.5 K that was far from $T_{ChSm}$. This dependence may be fitted well by the linear equation:

$$1/\lambda_m = X/\lambda_{mCOC} \tag{1}$$

Here, $\lambda_m \approx 355 \pm 3$ nm corresponds to pure COC($X$=1), and determination coefficient is 0.998.

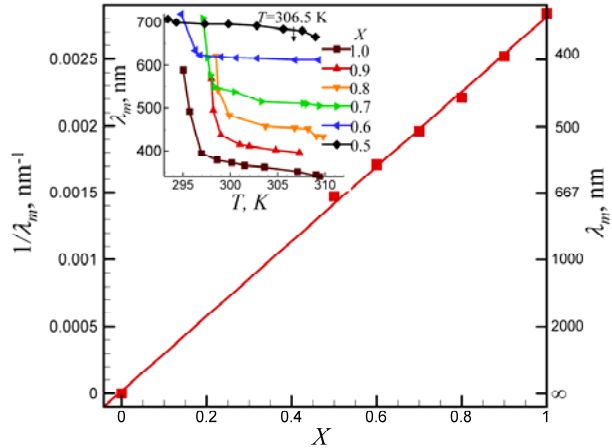

Figure 6. Reciprocal helical pitch $1/\lambda$ versus relative concentration X=COC/(COC+5CB) at T=306.5 K. The value of $\lambda_m$ was estimated from the maximum of selective reflection. Insert shows $\lambda_m$ versus temperature T for different values of X.

It is accepted that such type of linear dependence of the reciprocal pitch $1/\lambda_m$ versus concentration of optically active component $X$ is possible only for the case when molecules of nematic and cholesteric LCs are chemically similar (or almost similar) [24]. However, in our case, the chemical structures of COC and 5CB molecules are quite different.

Insert in Fig. 6 shows temperature dependencies of $\lambda_m$ at different values of *X*. Helical pitch is mainly untwisted when the value of *X* decreases (the value of $\lambda_m$ corresponds to the near-IR range at *X*<0.5), or temperature becomes closer to $T_{ChSm}$. In the vicinity of $T_{ChSm}$, the pitch exhibited a critical divergence, i.e., the power law behaviour typical for the smectic-A transition, where ν is the scaling exponent [25].

$$\lambda_m \propto (T - T_{ChSm})^{-\nu} \qquad (2)$$

This singularity in the pre-transition region can be explained by formation of large smectic clusters that cause a perturbation of director [26].

**Electrical conductivity of COC+NTs and COC+5CB+NTs composites**

The electrical conductivity $\sigma$ of COC + NTs composites was highly dependent on thermal pre-history of the samples, and this tendency was the most evident at large concentrations of NTs. Figure 7 presents examples of the temperature dependences of electrical conductivity $\sigma$ during the multiple heating-cooling cycles at *C*= 2 %. In these experiments, the samples were initially incubated at 255 K during 2-3 days in order to insure crystallization. After that, the temperature was growing at the constant rate of 2 K/min from 255 K to ≈320 K and then started to decrease up to 255 K with the same constant rate of 2 K/min. The heating-cooling cycles were repeated 3-5 times.

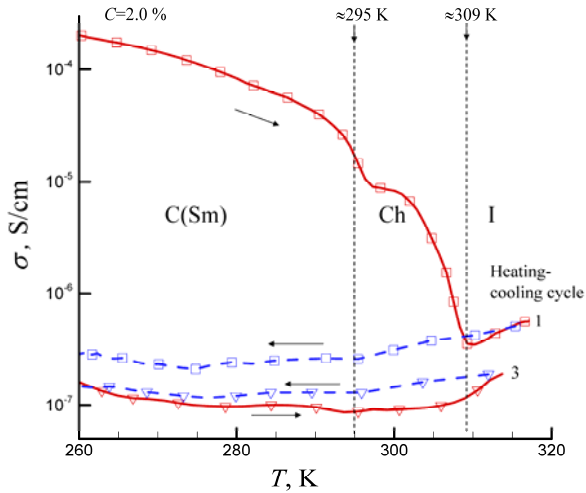

Figure 7. Electrical conductivity σ versus temperature T for heating and cooling cycles at NT concentration of C=2% in pure COC.

The most significant temperature hysteresis of $\sigma$ was observed in the first heating-cooling cycle. On first heating within the ranges of crystal (C) and cholesteric (Ch) phases, a rather abrupt decrease of electrical conductivity and evident inflections near the phase transition points at ≈295 K and ≈309 K were observed. Significant (by ≈3 orders of magnitude) decrease in electrical conductivity (Fig. 7) evidenced the presence of the effect of positive temperature coefficient (PTC) of resistivity, discussed in detail in our previous publications [8,27]. The similar PTC effect was observed earlier for ultrahigh molecular weight polyethylene [27] and LC [8] filled by NTs in the vicinity of the percolation threshold. It was explained by the damage of percolation networks due to thermally-induced expansion of the host matrix. However, the temperature hysteresis of $\sigma$ diminished in the sequential heating-cooling cycle and the $\sigma(T)$ curves became practically repetitive after the 2nd-3rd cycles for the temperature scanning within the isotropic, cholesteric and supercooled smectic A phases.

Figure 8a presents the temperature dependencies of $\sigma(T)$ for the cooling 3rd cycles at different concentrations *C* of NTs. The isotropic-cholesteric and cholesteric-smectic A phase transitions are practically indistinguishable on the obtained $\sigma(T)$ curves. From another side, the most pronounced temperature dependence of electrical conductivity was observed for pure COC. This dependence was close to the Arrhenius plot with activation energy of 22±3 kJ/mol. Moreover, the increase of NTs concentration resulted in initial decrease (within 0-0.3 %) and subsequent increase (at C> 0.3 %) of electrical conductivity. Such behavior may evidence formation of isolating solvation layers of COC that cover the NTs and prevent direct contacts between them at $C<0.3\%$. The similar effect of formation of isolating solvation layers was previously reported for aqueous suspensions of graphite stabilized by surface active substance Triton X305 [28].

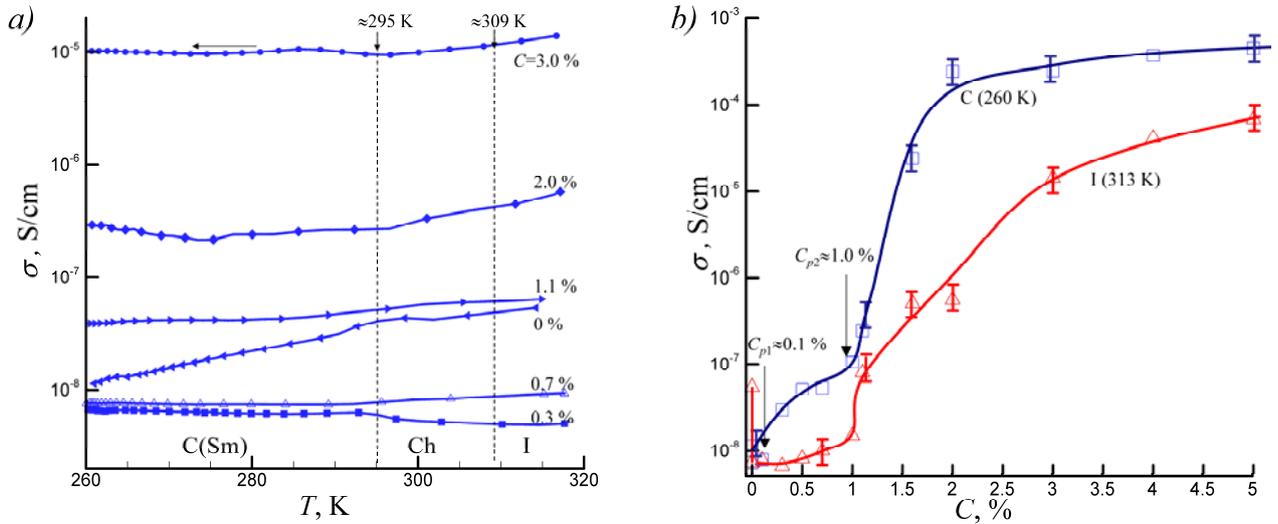

Figure 8. Electrical conductivity σ versus temperature T for the cooling cycles (3rd cycle) at different concentration C of NTs (*a*) and σ versus C at T=260 K (in crystalline phase, 1rt cycle) and at T=313 K (in isotropic phase) (*b*).

Moreover, introduction of NTs may result in binding of ionic charges at impurities that may control the conductivity mechanism in COC. Figure 8b presents concentration dependencies of σ(C) at two different temperatures: $T$=313 K (in isotropic phase) and $T$=260 K (in crystalline phase, 1$^{rt}$ cycle). In the isotropic phase the value of $\sigma$ passes through the minimum and continuously increases at $C \geq 1.0\%$. Its behaviour in the isotropic phase presumably reflects formation of conductive percolation networks. In the crystalline phase, the value of $\sigma$ continuously grows, and a fuzzy type percolation with multiple thresholds (two at least at $C \approx 0.3\%$ and $C \approx 1.0\%$, Fig. 8*b*) was observed in the studied COC+NTs composites. A similar fuzzy type percolation with multiple thresholds was previously observed for the smectogenic liquid crystal BBBA doped by NTs[23].

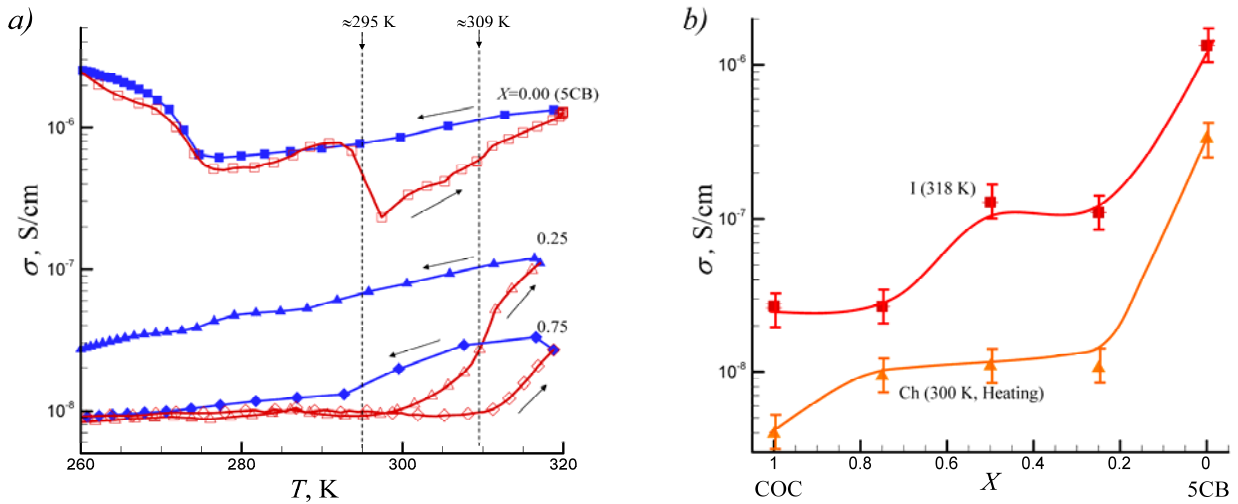

Figure 9. Electrical conductivity σ versus temperature T for heating-cooling cycles at different relative concentration X=COC/(COC+5CB) (*a*) and σ versus X at T=300 K (in cholesteric phase) and T=318 K (in isotropic phase) (*b*). The concentration of NTs C was 0.1 %.

Comparison of electrical conductivity behaviour of composites with various ratio of COC and 5CB components and fixed content of NTs (C=0.1 %) also shows clearly the presence of the percolation transition. Examples of σ(T) heating-cooling hysteresis for one and the same concentration of NTs, $C$=0.1% and different values of $X$ are presented in Fig. 9a. It is interesting that cooling $\sigma(T)$ curves always go over the heating $\sigma(T)$ curves in the whole range of the relative concentrations $X$=0.0-1.0. Introduction of even a small quantity of 5CB in the mixture (e.g., at large content of COC in the mixture, $X$<0.9) resulted in pronounced hysteretic behaviour during several ($3^{th}$ -$5^{th}$) heating-cooling cycles.

The hysteretic behaviour reflected the strengthening of electric contacts between different NTs owing to their intensive Brownian motions at elevated temperature in the isotropic phase. Figure 9b presents electrical conductivity $\sigma$ versus $X$ dependencies at $C$=0.1 % at two different temperatures: $T$=318 K (in the isotropic phase) and $T$=300 K (in the cholesteric phase, $3^{rt}$ cycle). Variation of the content of LC mixture allowed regulation of $\sigma$ within the two orders of magnitude. The most pronounced increase of $\sigma$ at X<0.2 reflected formation of large clusters of NTs promoted by "bad" 5CB solvent.

## 4. CONCLUSION

The main results of our findings can be summarized as follows:

- The good dispersing ability of the "good" cholesteric solvent COC and high stability of dispersion of carbon NTs in this solvent was demonstrated. The aggregation of NTs was unessential and hardly observable in this solvent, resulting in a rather small electrical conductivity σ of the COC+NTs composite. The fuzzy type percolation with multiple thresholds (two at least, at C≈0.3% and C≈1.0%) was observed. Suspensions of NTs in such solvent have demonstrated the presence of strong pre-history effects of phase transitions and temperature dependences of σ.

- Application of the mixtures of "good" (COC) and "bad" (5CB) nematic solvents provided the excellent tool for regulation of dispersion, stability and electrical conductivity of LC+NTs composites. Addition of 5CB in COC initiates pronounced heating-cooling hysteresis of electrical conductivity and appearance of clusters that can create the percolation network at the certain threshold value of $X$, which depends on the concentration of NTs in dispersion.

- The mixtures of COC (as well as other cholesterol esters) and 5CB are promising as good functional media with controllable and useful chiral and electrophysical properties.

## ACKNOWLEDGEMENTS

This work was supported by Projects 2.16.1.4 and 2.16.1.7 NAS of Ukraine and ITSU Project 4687.

## REFERENCES

[1] J. P. F. Lagerwall and G. Scalia, "A new era for liquid crystal research: Applications of liquid crystals in soft matter nano-, bio- and microtechnology," *Current Applied Physics* **12**(6), 1387–1412 (2012).
[2] H. Atkuri, G. Cook, D. R. Evans, C.-I. Cheon, A. Glushchenko, V. Reshetnyak, Y. Reznikov, J. West, and K. Zhang, "Preparation of ferroelectric nanoparticles for their use in liquid crystalline colloids," *J. Opt. A: Pure Appl. Opt.* **11**, 024006 (5pp) (2009).
[3] J. C. Payne, "Nanoparticle-chiral nematic liquid crystal composites," Massachusetts Institute of Technology (2006).
[4] P. Malik, A. Chaudhary, R. Mehra, and K. K. Raina, "Electro-optic, thermo-optic and dielectric responses of multiwalled carbon nanotube doped ferroelectric liquid crystal thin films," *Journal of Molecular Liquids* **165**, 7–11 (2012).
[5] R. Basu and G. S. Iannacchione, "Carbon nanotube dispersed liquid crystal: A nano electromechanical system," *Applied Physics Letters* **93**(18), 183105 (3 pages) (2008).
[6] L. Dolgov, O. Kovalchuk, N. Lebovka, S. Tomylko, and O. Yaroshchuk, "Carbon nanotubes," J. M. Marulanda, Ed., pp. 451–484, InTech (2010).

[7] N. Lebovka, T. Dadakova, L. Lysetskiy, O. Melezhyk, G. Puchkovska, T. Gavrilko, J. Baran, and M. Drozd, "Phase transitions, intermolecular interactions and electrical conductivity behavior in carbon multiwalled nanotubes/nematic liquid crystal composites," *Journal of Molecular Structure* **887**(1-3), 135–143 (2008).

[8] A. I. Goncharuk, N. I. Lebovka, L. N. Lisetski, and S. S. Minenko, "Aggregation, percolation and phase transitions in nematic liquid crystal EBBA doped with carbon nanotubes," *Journal of Physics D: Applied Physics* **42**(16), 165411 (2009).

[9] L. N. Lisetski, S. S. Minenko, A. V. Zhukov, P. P. Shtifanyuk, and N. I. Lebovka, "Dispersions of carbon nanotubes in cholesteric liquid crystals," *Molecular Crystals and Liquid Crystals* **510**, 43–50 (2009).

[10] S. Schymura and J. Lagerwall, "Carbon nanoparticles in cholesteric liquid crystals," in *Deutschen Bunsen-Gesellschaft e.V. 37. Arbeitstagung Flussigkristalle (2009, Stuttgart)* (2009).

[11] O. Koysal, "Conductivity and dielectric properties of cholesteric liquid crystal doped with single wall carbon nanotube," *Synthetic Metals* **160**(9-10), 1097–1100 (2010).

[12] C.-K. Chang, S.-W. Chiu, H.-L. Kuo, and K.-T. Tang, "Cholesteric liquid crystal-carbon nanotube hybrid architectures for gas detection," *Applied Physics Letters* **100**(4), 43501, AIP (2012).

[13] A. V. Melezhik, Y. I. Sementsov, and V. V. Yanchenko, "Synthesis of fine carbon nanotubes on coprecipitated metal oxide catalysts," *Russian Journal of Applied Chemistry* **78**(6), 917–923 (2005).

[14] M. Born and E. Wolf, *Principle of optics*, Pergamon, New York (1999).

[15] V. G. Denisenko, R. I. Egorov, and M. S. Soskin, "Stokes polarimetry in singular optics," *Proc. SPIE* **5477**, 41–46 (2004).

[16] M. S. Soskin, V. G. Denisenko, and R. I. Egorov, "Singular Stokes-polarimetry as new technique for metrology and inspection of polarized speckle-fields," *Proc. SPIE* **54758**, 79–85 (2004).

[17] V. V. Ponevchinsky, A. I. Goncharuk, V. I. Vasil'ev, N. I. Lebovka, and M. S. Soskin, "Cluster self-organization of nanotubes in a nematic phase: The percolation behavior and appearance of optical singularities," *JETP Letters* **91**(5), 241–244 (2010).

[18] L. M. Blinov, *Structure and properties of liquid crystals*, Springer Science (2011).

[19] L. N. Lisetski, N. I. Lebovka, S. V. Naydenov, and M. S. Soskin, "Dispersions of multi-walled carbon nanotubes in liquid crystals: A physical picture of aggregation," *Journal of Molecular Liquids* **164**(1-2), 143–147 (2011).

[20] L. N. Lisetski, S. S. Minenko, A. P. Fedoryako, and N. I. Lebovka, "Dispersions of multiwalled carbon nanotubes in different nematic mesogens: The study of optical transmittance and electrical conductivity," *Physica E: Low-Dimensional Systems and Nanostructures* **41**(3), 431–435 (2009).

[21] L. N. Lisetski, S. S. Minenko, V. V Ponevchinsky, M. S. Soskin, A. I. Goncharuk, and N. I. Lebovka, "Microstructure and incubation processes in composite liquid crystalline material (5CB) filled with multi walled carbon nanotubes," *Materialwissenschaft und Werkstofftechnik* **42**(1), 5–14 (2011).

[22] M. Zapotocky, L. Ramos, P. Poulin, T. C. Lubensky, and D. A. Weitz, "Particle-stabilized defect gel in cholesteric liquid crystals," *Science* **283**(5399), 209–212 (1999).

[23] N. I. Lebovka, L. N. Lisetski, A. I. Goncharuk, S. S. Minenko, V. V Ponevchinsky, and M. S. Soskin, "Phase transitions in smectogenic liquid crystal 4-butoxybenzylidene-4'-butylaniline (BBBA) doped by multiwalled carbon nanotubes," *Phase Transitions* **xx**(xx), 1–14 (2013).

[24] G. Chilaya, "Induction of chirality in nematic phases," *Rev. Phys. Appl. (Paris)* **16**(5), 193–208 (1981).

[25] S.-H. Chen and J. J. Wu, "Divergence of cholesteric pitch near the smectic-a transition in some cholesteryl nonanoate binary mixtures," *Molecular Crystals and Liquid Crystals* **87**(3-4), 197–209 (1982).

[26] A. V. Tolmachev, A. V. Tishchenko, and L. N. Lisetskii, "Optical properties and structural ordering of the planar texture of a cholesteric liquid crystal," *Sov. Phys. JETP* **48**(2), 333–338 (1978).

[27] M. O. Lisunova, Y. P. Mamunya, N. I. Lebovka, and A. V. Melezhyk, "Percolation behaviour of ultrahigh molecular weight polyethylene/multi-walled carbon nanotubes composites," *Eur. Polym. J.* **43**, 949–958 (2007).

[28] V. Moraru, N. Lebovka, and D. Shevchenko, "Structural transitions in aqueous suspensions of natural graphite," *Colloids and Surfaces A: Physicochemical and Engineering Aspects* **242**(1-3), 181–187 (2004).